\documentclass[journal=jacsat,manuscript=article]{achemso}

\usepackage{times}
\usepackage{color}
\usepackage{graphicx}
\usepackage{textcomp}
\usepackage{amsmath,amsbsy,amsfonts,mathrsfs}

\newcommand{\be}{\begin{equation}}
\newcommand{\ee}{\end{equation}}
\newcommand{\bea}{\begin{eqnarray}}
\newcommand{\eea}{\end{eqnarray}}
\newcommand{\bg}{\begin{figure}}
\newcommand{\eg}{\end{figure}}
\newcommand{\bi}{\begin{itemize}}
\newcommand{\ei}{\end{itemize}}
\author{Juan I. Beltr\'an}
\email{juan.beltran@uam.es}
\affiliation[Universidad Aut\'onoma de Madrid]{Departamento de F\1sica Te\'orica de la Materia Condensada, 
\\ Universidad Aut\'onoma de Madrid, E-28049 Madrid, Spain}

\author{Fernando Flores}
\email{fernando.flores@uam.es}
\affiliation[Universidad Aut\'onoma de Madrid]{Departamento de F\1sica Te\'orica de la Materia Condensada, 
\\ Universidad Aut\'onoma de Madrid, E-28049 Madrid, Spain}

\author{Jos\'e I. Mart\1nez}
\affiliation[Universidad Aut\'onoma de Madrid]{Departamento de F\1sica Te\'orica de la Materia Condensada, 
\\ Universidad Aut\'onoma de Madrid, E-28049 Madrid, Spain}

\author{Jos\'e Ortega}
\affiliation[Universidad Aut\'onoma de Madrid]{Departamento de F\1sica Te\'orica de la Materia Condensada, 
\\ Universidad Aut\'onoma de Madrid, E-28049 Madrid, Spain}

\title[Energy Level Alignment in Organic-Organic Heterojunctions: The TTF-TCNQ Interface]
{Energy Level Alignment in Organic-Organic Heterojunctions: The TTF-TCNQ Interface}

\begin{document}

\newpage

\begin{abstract}
The energy level alignment of the two organic materials forming the TTF-TCNQ interface is analyzed by means of a local 
orbital DFT calculation, including an appropriate correction for the transport energy gaps associated with both materials. 
These energy gaps are determined by a combination of some experimental data and the results of our calculations for the 
difference between the TTF$_{HOMO}$ and the TCNQ$_{LUMO}$ levels. We find that the interface is metallic, as predicted 
by recent experiments, due to the overlap (and charge transfer) between the Density of States corresponding to these 
two levels, indicating that the main mechanism controlling the TTF-TCNQ energy level alignment is the charge transfer 
between the two materials. We find an induced interface dipole of 0.7 eV in good agreement with the experimental evidence. 
We have also analyzed the electronic properties of the TTF-TCNQ interface as a function of an external bias voltage 
$\Delta$ between the TCNQ and TTF crystals, finding a transition between metallic and insulator behavior for $\Delta\sim0.5$ eV. 
\end{abstract}


\section{Introduction}

\begin{figure}
\centerline{\includegraphics[width=\columnwidth]{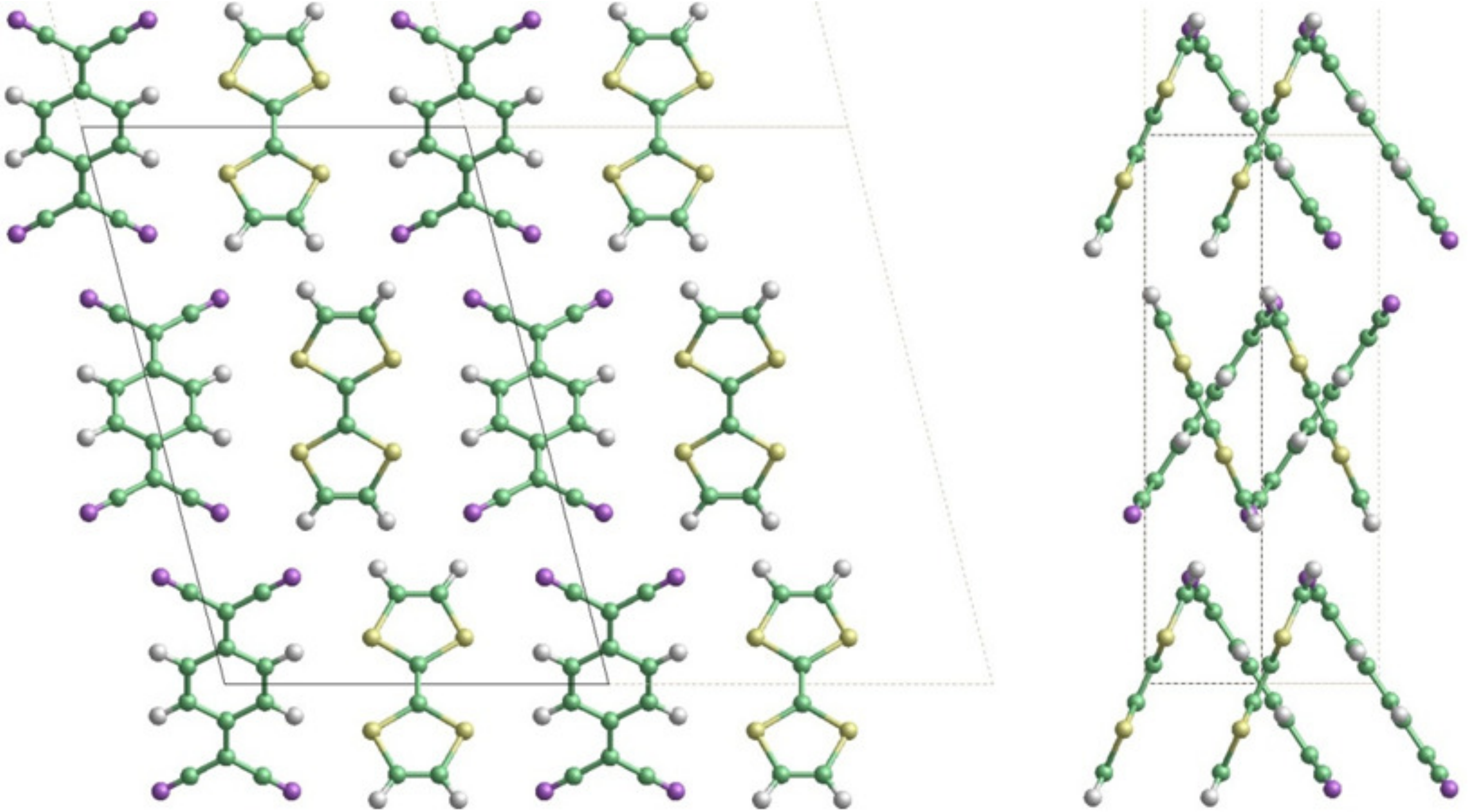}}
\smallskip \caption{(color online) TTF-TCNQ bulk structure with its view-plane along the {\bf b} 
and {\bf a} lattice vectors, which corresponds to $\hat{y}$ and $\hat{x}$ axis, for the left and 
right pictures, respectively.} \label{P1}
\end{figure}

Organic electronics is an area of important research effort where organic materials provide an immense variety of properties and atomic structures. 
In these materials, interfaces play a key role in fundamental phenomena (e.g. charge separation and charge recombination) for electronic applications 
such as solar cells and light emitting diodes, to name a few~\cite{1,2,3}. 

TTF/TCNQ (tetrathiofulvalene-7,7,8,8,-tetracyanoquinodimethane) is a very interesting organic bulk material that was already studied in the early 70's 
showing novel phenomena, like large e$^{-}$ conductivity and superconductivity~\cite{4}. TTF/TCNQ shows three phase transitions which have been related to 
condensed charge-density waves, due to Peierls distortions, with transition temperatures of 38, 49 and 54 K~\cite{4}. Also, the experimental band widths of 
TTF and TCNQ crystal-bulks are 0.95 and 2.5 eV respectively, while density functional theory (DFT) calculations based on local and semi-local exchange-correlation 
(XC) functional provide systematically, for TTF and TCNQ, band widths of around 0.65 and 0.70 eV respectively~\cite{5,6,7,8}. These intriguing properties have prompted a 
lot of experimental and theoretical studies on the properties of bulk TTF/TCNQ, see e.g.~\cite{5,6,7,8,9,10}; thin films of mixed TTF-TCNQ have also been studied on different 
substrates like graphene and Au(111)~\cite{11,12}. 

However, the large amount of work on bulk and thin film properties of TTF/TCNQ contrasts with the much lesser attention dedicated to the interface of the two 
organic materials. One of the few studies~\cite{13} has found that the interface retains its metallic character due to the charge transfer between the two molecules. 
The existence of an interface dipole layer ($IDL$)~\cite{14,15}, generated by charge transfer or/and by polarization effects, and its corresponding vacuum level shift 
($VLS$)~\cite{1}, shows that the Schottky-Mott limit is not satisfied at this interface. Careful spectroscopic characterization of TTF-TCNQ interfaces over different 
substrates shows a $VLS$ of 0.6 eV and a $\Delta_{HOMO}$ = HOMO$_{TTF}-$HOMO$_{TCNQ}$ of either 2.1 or 2.4 eV~\cite{16,17}.
 
In this communication we consider the interface between TTF and TCNQ crystals, and analyze the interface energy level alignment and barrier height formation, 
using a local orbital DFT-approach in which an appropriate correction is included to properly describe the transport gaps of both materials. In section II we 
present the interface geometries, and in section III we discuss how we determine the transport energy gaps for TTF and TCNQ and give some details about our DFT-calculations. 
In section IV, we present our results. In particular, our analysis shows that the TTF-TCNQ interface is metallic and that the energy difference between the TCNQ$_{LUMO}$ 
and TTF$_{HOMO}$ is pinned to around 0.8 eV. We have also analyzed how the interface properties depend on the application of an external voltage between the TCNQ and TTF 
crystals, finding a transition from a metallic to an insulating interface for a voltage of $\sim$ 0.5 eV. Finally, the interfacial geometry has also been analyzed by means 
of a DFT calculation combined with a semi-empirical parametrization of the van der Waals interaction.

\begin{figure}
\centerline{\includegraphics[width=\columnwidth]{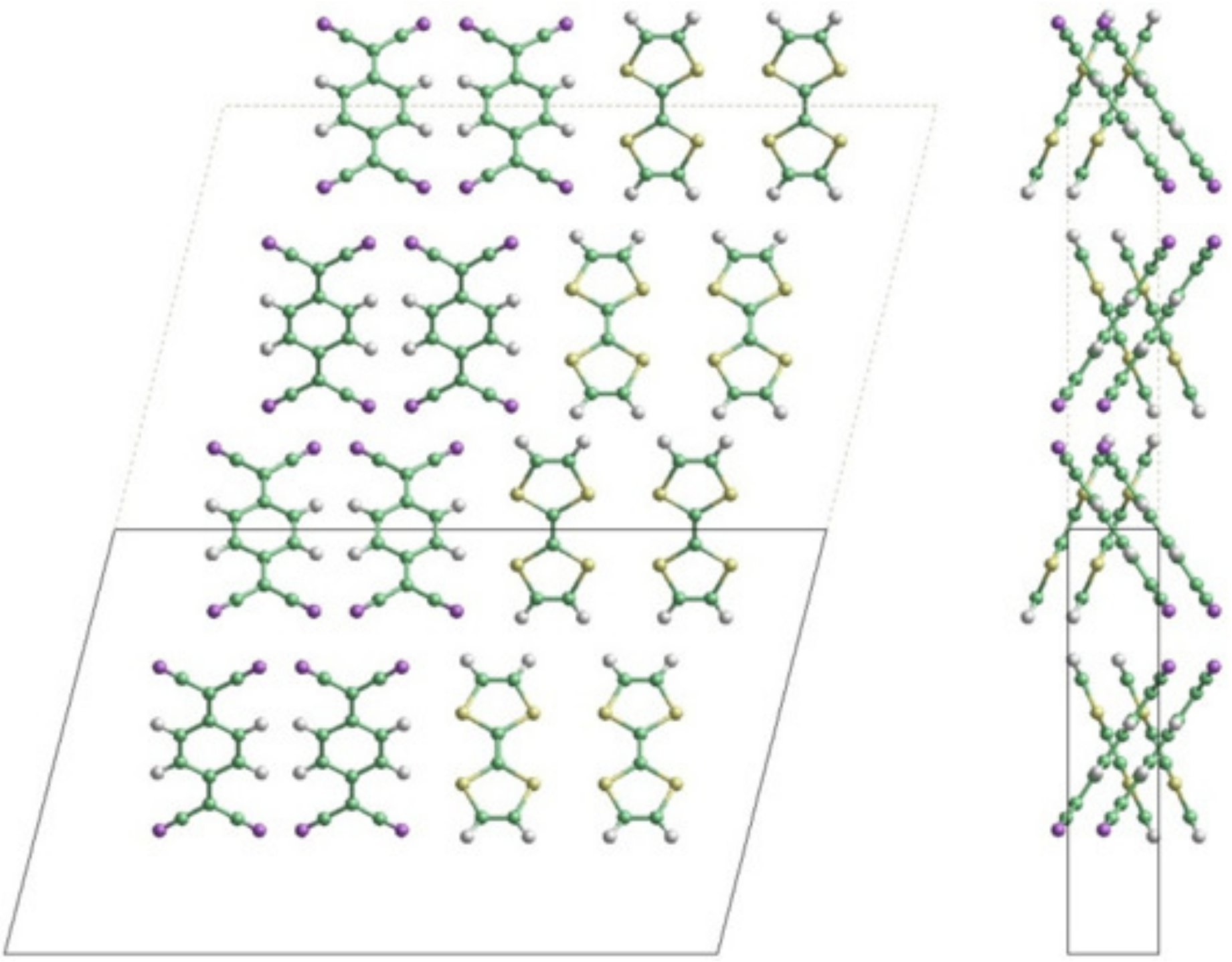}}
\smallskip \caption{(color online) Similar to~\ref{P1} but for the bilayer interface.} \label{P2}
\end{figure}

\section{Interface Geometry}

At room temperature, TTF/TCNQ bulk presents a monoclinic structure with lattice parameters {\bf a} = 12.298~\AA, {\bf b} = 3.819~\AA~and {\bf c} = 18.468~\AA, and an angle 
$\hat{\beta}$ = 104.46\textdegree~\cite{18} (see~\ref{P1}). However, detailed experimental information of both the geometry and the energetics at the interface is still missing 
due to inherent difficulties in its structural characterization. For the TTF-TCNQ surface, though, it has been measured by scanning tunneling microscopy~\cite{19} and angle-dependent 
near-edge X-ray adsorption fine structure~\cite{20} that the cleaved surfaces are highly ordered and retain the periodicity of the bulk. Thus, in this work we have considered the 
interfacial structure arising from selecting the interface geometry between TTF and TCNQ layers from the TTF/TCNQ bulk structure along the (100) direction (see~\ref{P2}). In 
the DFT calculations we have used bulk periodic boundary conditions along {\bf b} and {\bf c} vectors, while for the a direction we have considered two cases: a monolayer interface, 
with only one layer of each material, and a bilayer interface, with 2 TTF and TCNQ layers. This amounts to 2 TTF + 2 TCNQ or 4 TTF + 4 TCNQ molecules in the unit cell, respectively.
In the bilayer interface, the second layer is generated from the interfacial plane by adding parallel molecules which are shifted in a similar way as in their independent crystals for 
both the TTF and TCNQ bulk structures -- see Cambridge Structural Database -- and satisfying the periodic boundary conditions of keeping {\bf b} and {\bf c} vectors from the bulk structure 
(see~\ref{P2}).

\begin{figure}
\centerline{\includegraphics[width=6cm]{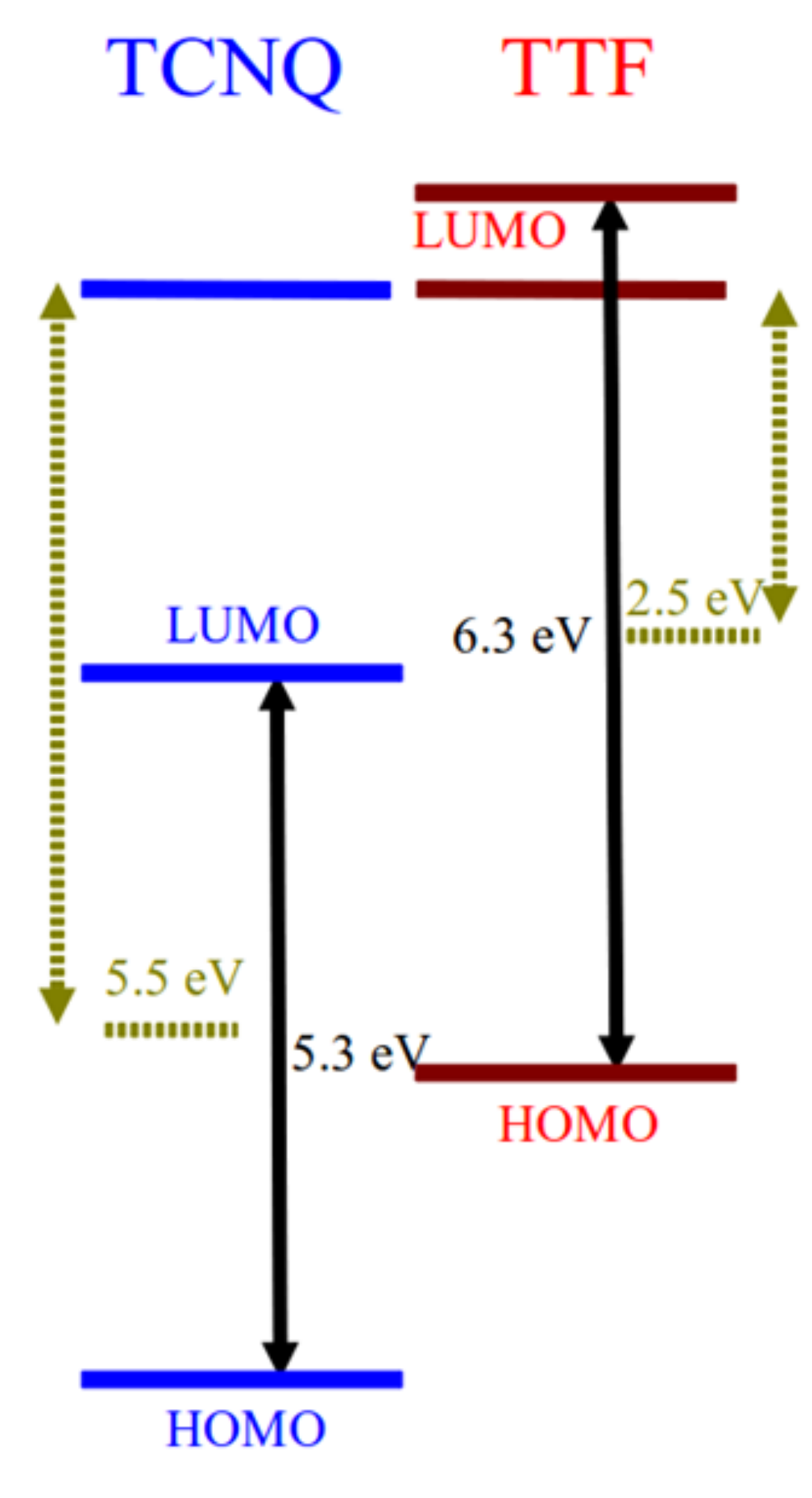}}
\smallskip \caption{(color online) Energy level diagram for the isolated TTF and TCNQ molecules 
obtained from $\Delta$-SCF calculations (explained in the main text).} \label{P3}
\end{figure}

\section{TTF-TCNQ Interface Energy Level Alignment}

\subsection{$\Delta$-SCF calculations}

As a first step in order to determine the energy level alignment at the TTF-TCNQ interface, we have performed $\Delta$-SCF calculations for the isolated TTF and TCNQ molecules. 
In these calculations, the HOMO and LUMO levels are determined from total energy DFT calculations~\cite{21,22} for molecules with $N$, $N+1$ and $N-1$ electrons ($N$ corresponding 
to the neutral molecule):
\bea \label{eq1}
\varepsilon_{HOMO}&=&E[N]-E[N-1], \nonumber \\
\varepsilon_{LUMO}&=&E[N+1]-E[N],
\eea where $E[N_{\alpha}]$ is the total energy for a molecule with $N_{\alpha}$ electrons. The results of these calculations are summarized in~\ref{P3}. We obtain a transport gap 
(difference between $\varepsilon_{LUMO}$ and $\varepsilon_{HOMO}$) of 5.3 eV and 6.3 eV for TCNQ and TTF, respectively, while the mid-gap positions are found to be located at 5.5 eV 
(TCNQ) or 2.5 eV (TTF) below the Vacuum Level ($VL$).

\begin{figure}
\centerline{\includegraphics[width=\columnwidth]{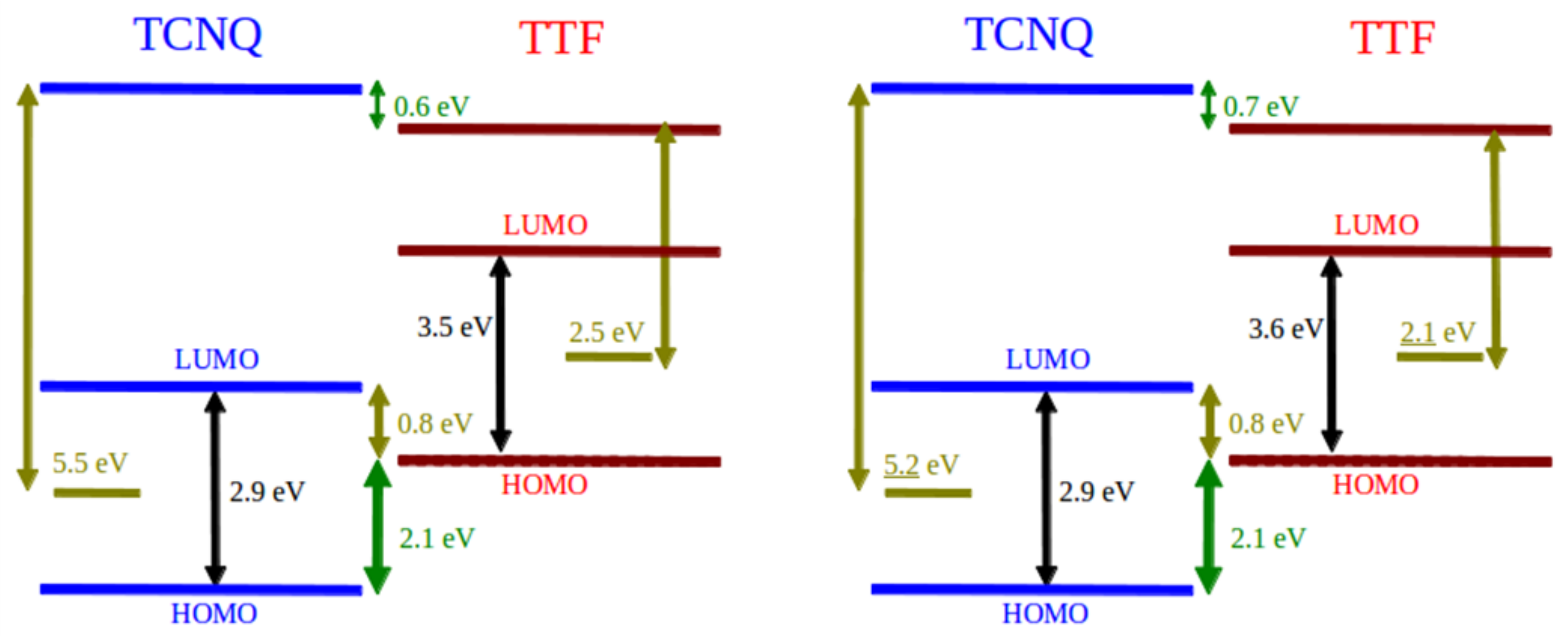}}
\smallskip \caption{(color online) Left panel: energy level diagram for the TCNQ-TTF interface 
obtained from a combination of experimental and theoretical information. Right panel: energy levels 
in the DFT calculation for the periodic interface shown in~\ref{P2}. In this calculation the 
{\it initial} transport gaps and mid-gap positions have been adjusted to reproduce those given in 
the left panel of this Figure (see main text).} \label{P4}
\end{figure}

\subsection{Energy levels at the interface}

The energy level values for the isolated molecules have to be corrected due to the interaction with the other molecules in the crystal and at the interface. The effect of this ‘environment’ 
is three-fold: (a) broadening of the energy levels into bands; (b) relative shift of the TTF and TCNQ molecular levels due to electrostatic and Pauli exclusion effects; (c) Polarization or 
screening effects, shifting in different directions occupied and empty states. Point (b) gives rise to a Vacuum Level Shift ($VLS$), or interface potential, between both crystals, while the 
effect of point (c) is a reduction of the transport gaps at the interface (or in the crystals) as compared with the values for isolated TTF or TCNQ molecules, $E_{g}^{T} = (E_{g}^{T})^{0} 
- \delta U$. For example, for a molecule interacting with a metal, image potential effects reduce the transport gap as follows: $E_{g}^{T} = (E_{g}^{T})^{0} - eV_{IM}$, where $V_{IM}$ 
is the image potential~\cite{23}. In the case of an organic crystal (or interface) a similar effect is present, due to the screening/polarization of the other molecules.
 
In order to obtain the energy level alignment for the TTF-TCNQ interface, we have used a combination of experimental and theoretical information:
\bi
\item[(1)] The $VLS$ at the interface is 0.6 eV, as determined experimentally~\cite{17}. In these experiments it is also found that $\Delta\varepsilon_{HOMO} = \varepsilon_{HOMO}(TTF)
- \varepsilon_{HOMO}(TCNQ) =$ 2.1 eV~\cite{17}.
\item[(2)] The mid-gap energies obtained in the $\Delta$-SCF calculations mentioned above are 5.5 eV (TCNQ) and 2.5 eV (TTF) below their corresponding Vacuum Level ($VL$) positions.
\item[(3)] As discussed below, we find that $\varepsilon_{LUMO}(TCNQ) - \varepsilon_{HOMO}(TTF) \approx$ 0.8 eV, a result that is related to the experimental observation that the interface 
is metallic~\cite{14}, see below.
\ei

Using this information, we can deduce that at the interface the transport gap is 2.9 eV for TCNQ, and 3.5 eV for TTF, i.e. screening/polarization effects reduce the transport gaps by 
2.4 eV (TCNQ) and 2.8 eV (TTF). \ref{P4} (left panel) summarizes the resulting energy level alignment for the TCNQ-TTF interface. 

\begin{figure}
\centerline{\includegraphics[width=10cm]{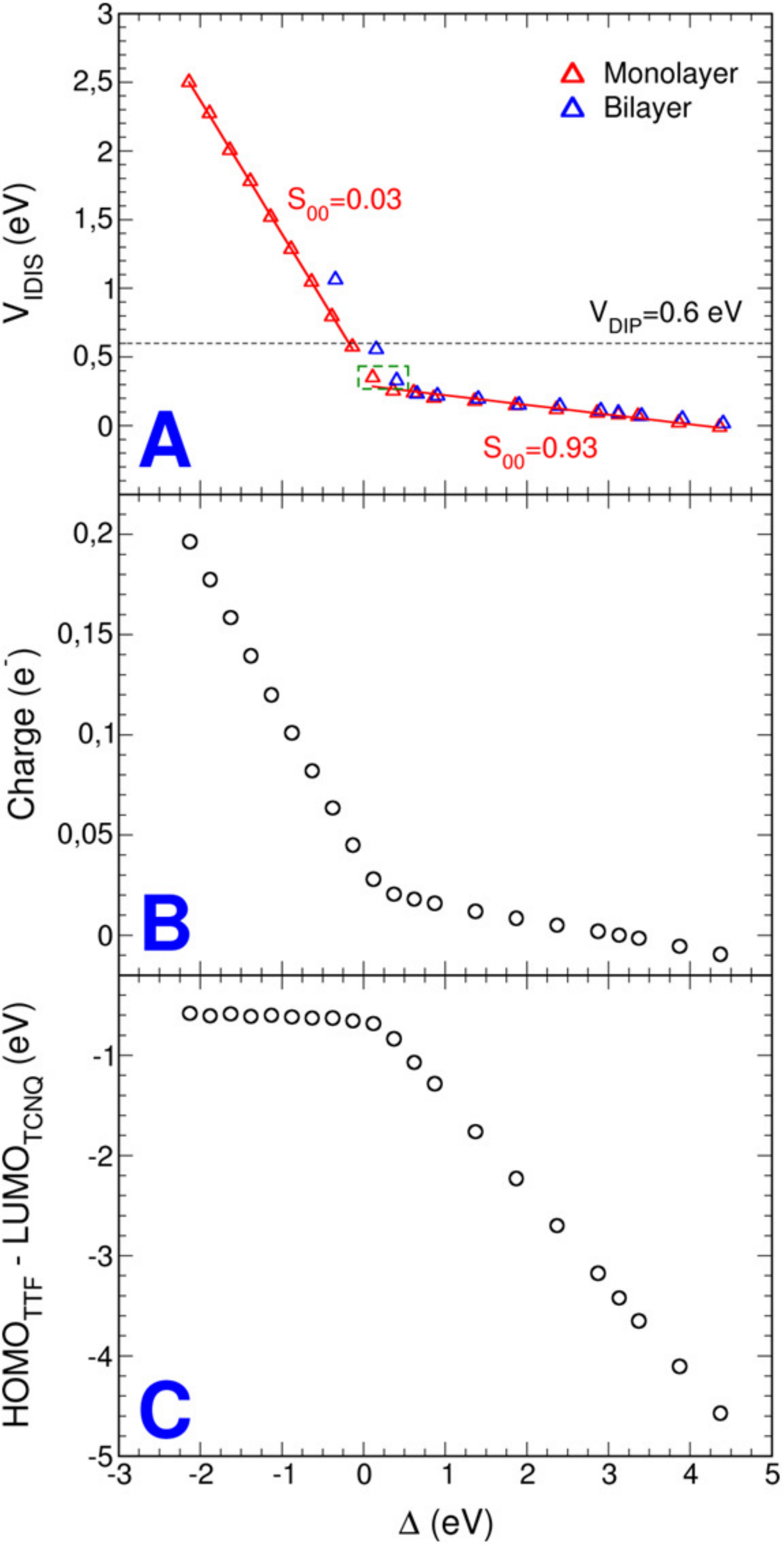}}
\smallskip \caption{(color online) A) V$_{IDIS}$ vs. $\Delta$ (initial shift of the TCNQ levels w.r.t.
the TTF levels, see text) for the monolayer and bilayer interfaces;  $\Delta$ = 0 corresponds to the 
realistic interface configuration. The experimental $VLS$ (or V$_{IDIS}$) value, 0.6 eV, is included 
as a dashed horizontal line; the screening parameter for each of the two regimes is also shown; 
B) Transfer of charge vs. $\Delta$ for the monolayer interface; and C) Value of [$\varepsilon_{HOMO}$(TTF)$-
\varepsilon_{LUMO}$(TCNQ)] vs. $\Delta$ for the monolayer interface.} \label{P5}
\end{figure}

\subsection{DFT calculations}

We have performed DFT calculations for the monolayer and bilayer periodic interface structures presented in section II (see~\ref{P2}). In these calculations we have used a local-orbital 
DFT code~\cite{24}, and the initial transport gaps for TCNQ and TTF have been adjusted to the values given in~\ref{P4} (left panel) using a scissor operator; besides, the initial TTF and TCNQ 
molecular levels are shifted in such a way that their initial $VL$ positions have the same value, and the initial mid-gap positions correspond to the values given in~\ref{P4} (left panel) 
(see e.g. Refs.~\cite{25,26,27} and references therein for details). Other technical details are the use of the local density approximation (LDA) and an optimized basis set of $s$ (H) and 
$sp^{3}d^{5}$ (C, N and S) Numerical Atomic-like Orbitals~\cite{28}, with the following cut-off radii (in a.u.): $s$=4.1 for H, and $s$=4.0, 3.6, 4.2, $p$=4.5, 4.1, 4.7, and $d$=5.4, 5.2, 5.5 
for C, N and S, respectively. 

\begin{figure}
\centerline{\includegraphics[width=\columnwidth]{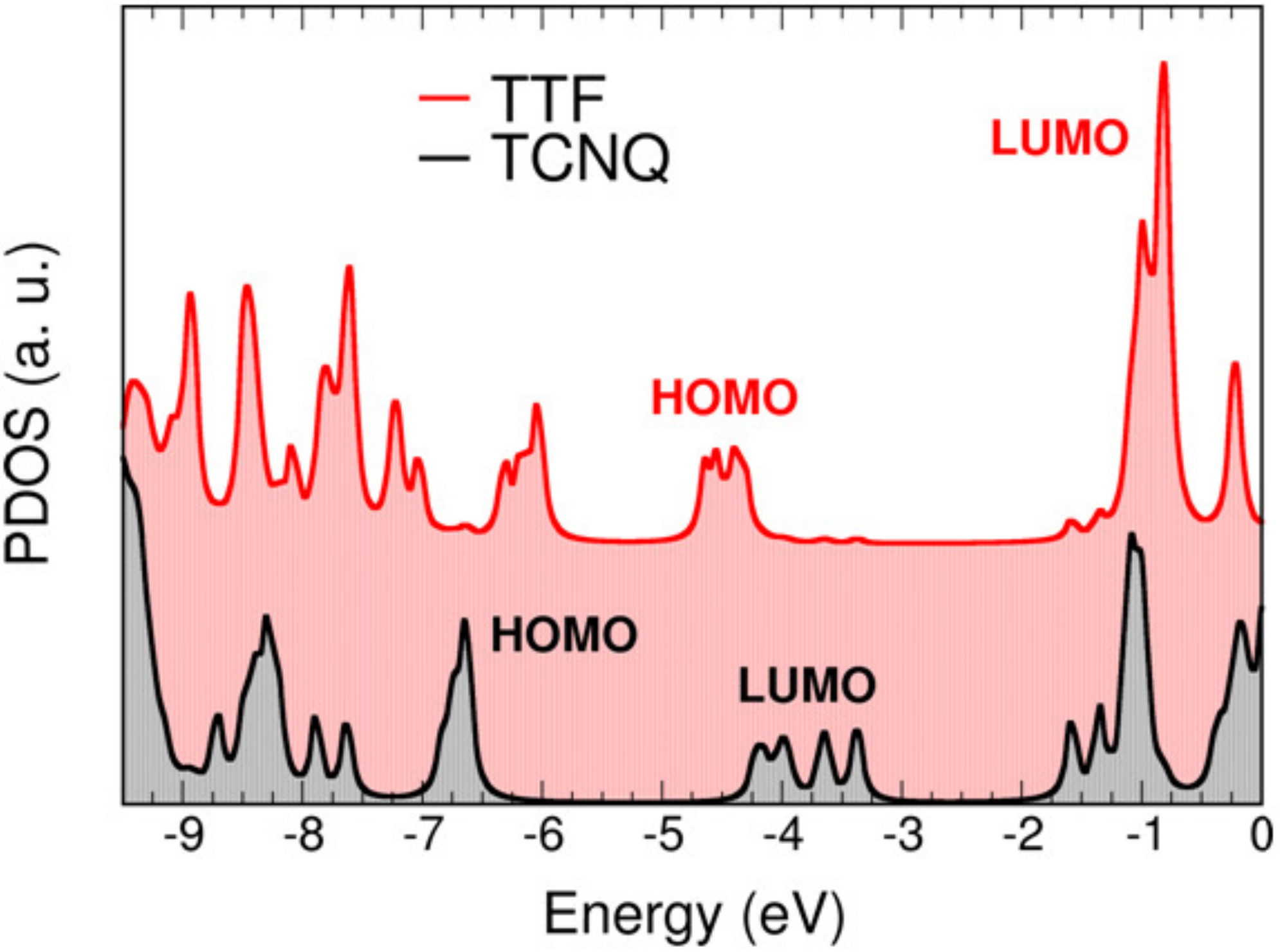}}
\smallskip \caption{(color online) Density of states projected on either the TCNQ or TTF of the 
monolayer interface for $\Delta$ = 0.} \label{P6}
\end{figure}

\section{Results and Discussion}

The right panel of~\ref{P4} shows the TCNQ-TTF interface energy level diagram obtained in our DFT calculations (explained in third subsection of previous section) with the initial 
conditions as explained above. This self-consistent result is quite similar to the energy level diagram obtained in second subsection of previous section, shown in the left panel 
of~\ref{P4}. The band-gaps and interface dipole are quite similar in both panels of~\ref{P4}, however the mid-gap positions move slightly upwards by around 0.3-0.4 eV. This shift 
is due to the Pauli repulsion (related to the Pauli exclusion principle).

\begin{figure}
\centerline{\includegraphics[width=\columnwidth]{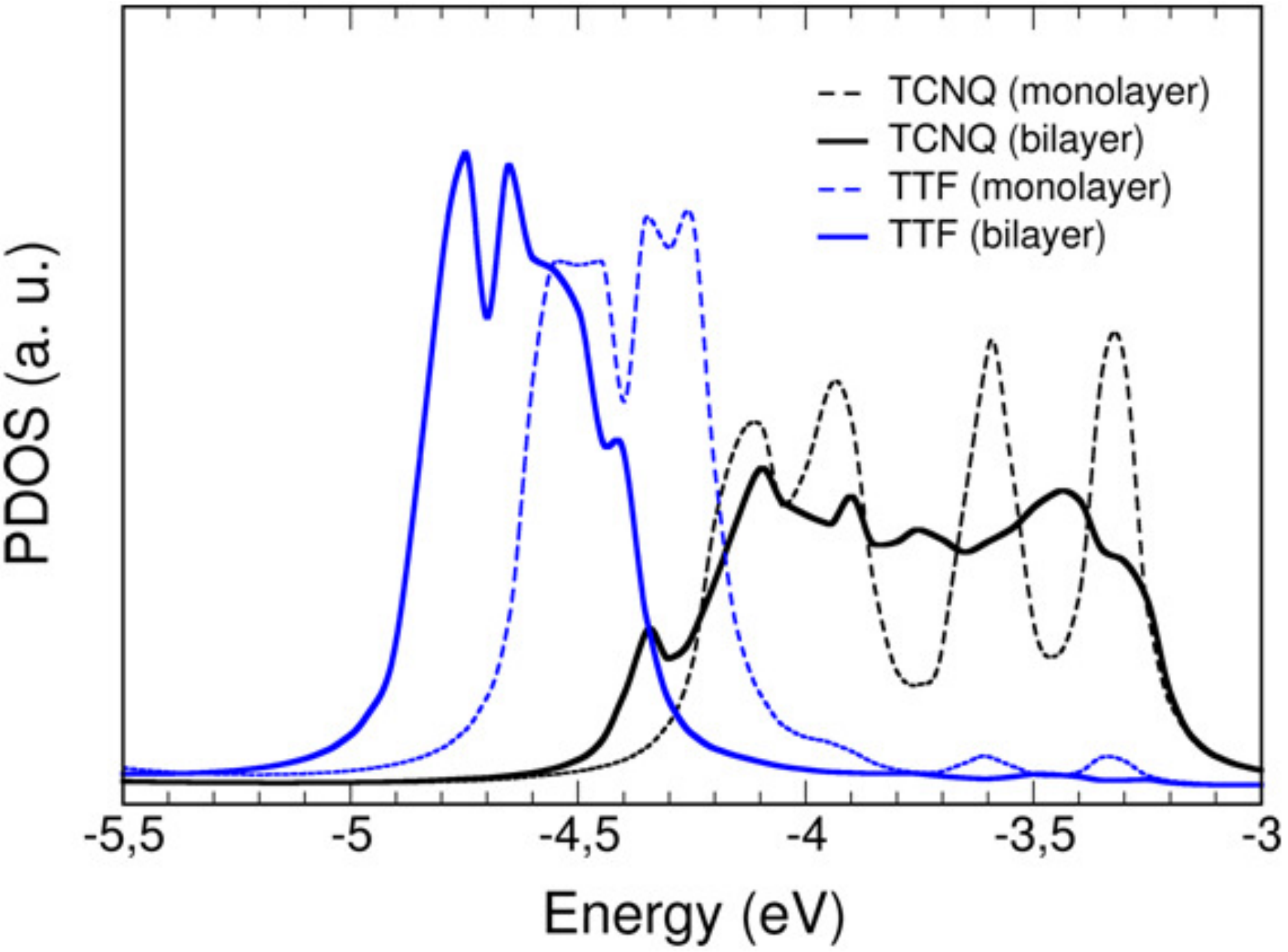}}
\smallskip \caption{(color online) DOS of the monolayer and bilayer interfaces projected on each 
material for $\Delta$ = 0. The PDOS intensity of the monolayer interface has been scaled according 
with the number of atoms included in the bilayer case.} \label{P7}
\end{figure}

In order to analyze the mechanism responsible for the energy level alignment at the TTF-TCNQ interface, we have performed DFT calculations as described in third subsection of previous section, 
in which we shift by $\Delta$ the initial (before self-consistency) position of the molecular levels of TCNQ with respect to the molecular levels of TTF;  $\Delta$ = 0 corresponds to equal 
initial vacuum level positions, and $\Delta>$ 0 corresponds to a destabilization of the TCNQ levels, w.r.t. the TTF levels. This shift simulates the application of a bias potential of 
value $\Delta$ between the two crystals. \ref{P4} (right panel) represents the final levels (after self-consistency) for the $\Delta$ = 0 case. A change $\Delta$ in the initial relative 
position of the TCNQ and TTF energy levels will influence the induced dipole and the rest of the electronic properties at the interface. The final $VLS$ will depend on the initial misalignment 
setup $\Delta$ (e.g. the external bias between the two crystals) and on the induced potential after electronic self-consistency, $V_{IDIS}$, in such a way:
\be \label{eq2}
VLS = \Delta + V_{IDIS}.
\ee In~\ref{P5}A we depict $V_{IDIS}$ as a function of $\Delta$ for the mono and bilayer-interfaces. In both graphs two different regimes are clearly observed, split by a crossing point 
located at $\Delta$ = 0.2 (0.45) eV for the monolayer (bilayer) case. For both regimes, we see that:
\be \label{eq3}
\delta V_{IDIS} = -(1-S)\delta\Delta,
\ee $S$ being a screening parameter~\cite{23}, taking the value 0.03 for $\Delta<$ 0.2 (0.45) eV and 0.93 for $\Delta >$ 0.2 (0.45) eV. The case of $\Delta<$ 0.2(0.45) eV corresponds to a 
metallic phase in which the $\varepsilon_{LUMO}(TCNQ)$ and $\varepsilon_{HOMO}(TTF)$ Densities of States (DOS) are overlapping, as shown in ~\ref{P6} and~\ref{P7}; in this case, the 
difference between  $\varepsilon_{LUMO}(TCNQ)$ and $\varepsilon_{HOMO}(TTF)$ is almost constant ($\approx$ 0.8 eV), due to the transfer of charge between these two levels. 
For the other phase, $\Delta >$ 0.2(0.45) eV, we find a typical heterojunction with small screening and a small induced potential, $V_{IDIS}$. 

This is illustrated in~\ref{P5}B-C, where we depict, for the monolayer interface, the charge transfer (positive if electron charge flows from TTF to TCNQ) and the value of 
$[\varepsilon_{HOMO}(TTF) - \varepsilon_{LUMO}(TCNQ)]$ vs. $\Delta$. For both plots the crossing point between the two different regimes is located at the same position than 
for $V_{IDIS}$, close to $\Delta =$ 0.2 eV, dividing the charge transfer plot into a regime with a large increase of charge transfer as $\Delta$ is reduced, for $\Delta <$ 0.2 eV, 
and another in which the transfer of charge is small and changes rather slowly as $\Delta$ is increased, for $\Delta >$ 0.2 eV. In~\ref{P5}C we see that for $\Delta >$ 0.2 eV the 
value of $[\varepsilon_{HOMO}(TTF) - \varepsilon_{LUMO}(TCNQ)]$ vs. $\Delta$ presents a slope close to -1, while for $\Delta <$ 0.2 eV $[\varepsilon_{HOMO}(TTF) - \varepsilon_{LUMO}(TCNQ)]$
is approximately constant, with a value of -(0.7-0.8) eV. This value ($\varepsilon_{HOMO}(TTF) - \varepsilon_{LUMO}(TCNQ)\approx$ 0.8 eV) has been used in second subsection of previous section 
(see left panel of~\ref{P4}) to deduce the TCNQ and TTF energy level positions at the interface. It is worth mentioning that zero charge transfer corresponds to the case of $\Delta =$ 3.12 eV; 
for this alignment $V_{IDIS}$ is not, however, zero, but takes the values 0.10 and 0.11 eV for the mono and the bilayer cases, respectively. These induced potentials are created by the 
polarization effects of the molecule.
  
In~\ref{P5} we observe that from the monolayer to the bilayer case the crossing point moves around 0.25 eV to positive values of $\Delta$. This is explained in~\ref{P7}, that shows, 
for the monolayer and bilayer interfaces, the DOS projected on each material for a value of $\Delta$ close to the crossing point. We notice that there is a broadening for the bilayer 
case of the $LUMO(TCNQ)$ DOS, which shifts by 0.25 eV the position where this DOS starts to overlap with the DOS corresponding to the $HOMO(TTF)$, as compared with the monolayer case.

\begin{figure}
\centerline{\includegraphics[width=\columnwidth]{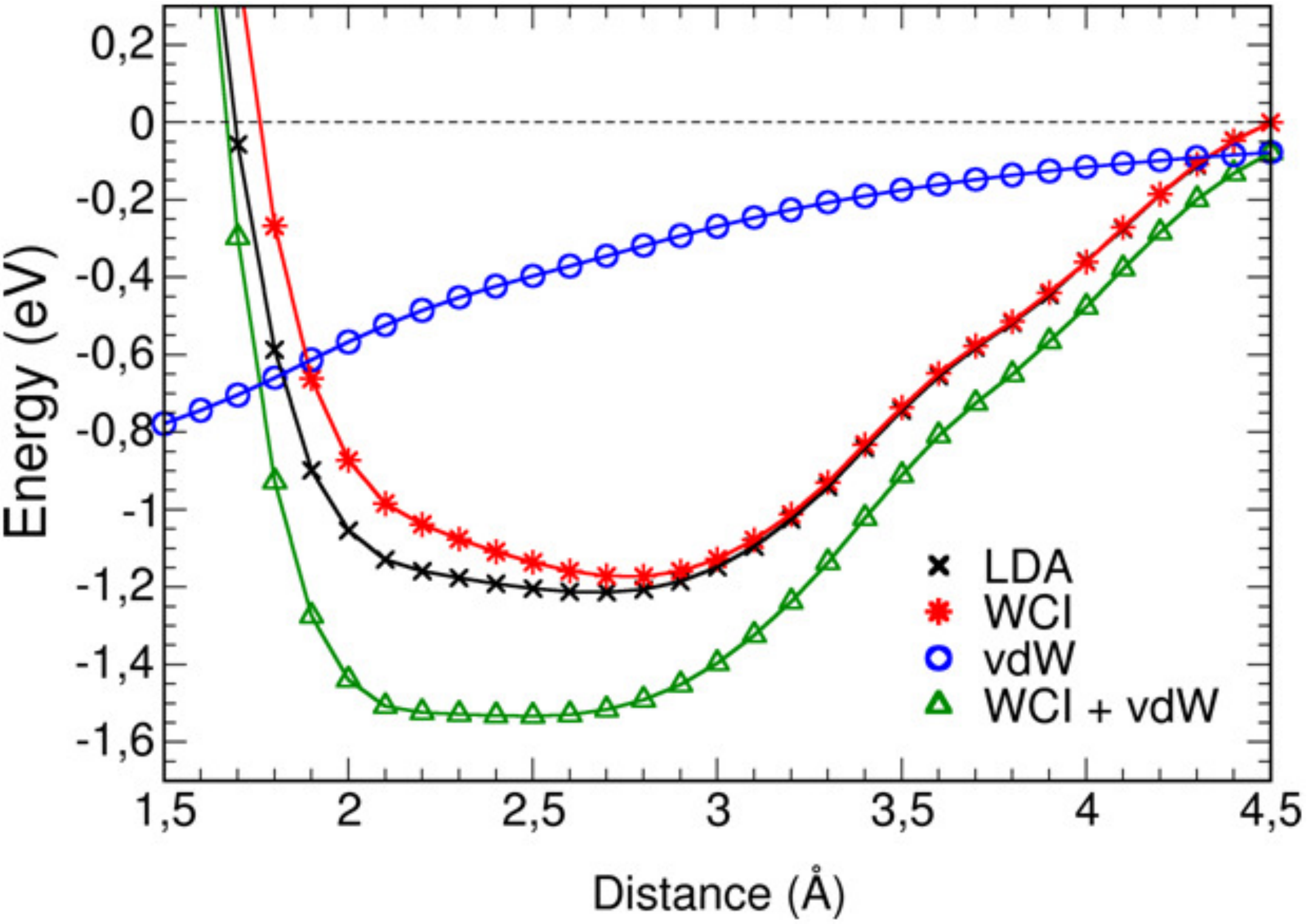}}
\smallskip \caption{(color online) Energy of the monolayer-interface as a function of the distance 
between the two planes.} \label{P8}
\end{figure}

Finally, as a check of the interface geometry used (see~\ref{P2}), we have calculated the total energy as a function of the distance between the TTF and TCNQ layers at the interface. 
\ref{P8} shows the total energy curves for the LDA calculation and the van der Waals (vdW) interactions energy; this energy is included in a semiempirical fashion as an atractive 
$-f_{D}(R)C_{6}/R^{6}$ atom-atom term. $f_{D}(R)$ is the Grimme damping expresion~\cite{29} and $C_{6}$=0.7$\times C_{6}'$, where $C_{6}'$ is the interatomic parameter calculated using 
the London dispersion relation as a function of the atomic polarizability and the first ionization potential~\cite{30}; 0.7 is a constant correction calculated from a detailed analysis 
of the C--C interaction~\cite{31}, which is applied to every $C_{6}'$ parameter. In order to prevent for the over-counting that would appear including both, the exchange-correlation energy 
provided by a conventional LDA and the correlation energy associated with the long-range vdW interaction, we have also performed a ``corrected LDA'' calculation in which the 
exchange-correlation energy for the complete system is calculated as the sum of the exchange-correlation energies for each subsystem, taken each one independently, neglecting in this 
way the effect of the overlapping densities in the exchange-correlation energy~\cite{32}. In~\ref{P8} this interaction energy is labeled weak chemical interaction (WCI). The minimum of 
the WCI+vdW total energy curve is around 2.5~\AA, which is close to the experimental value for the distance between TTF and TCNQ layers in bulk TTF/TCNQ; notice, however, the very 
flat energy profile (within 0.02 eV) from 2.2 to 2.7~\AA. This result suggests that starting from the bulk structure to generate the interface geometry is a good approximation.

\section{Conclusions}

In conclusion, we have presented a DFT analysis of the energy level alignment in the TTF/TCNQ interface, introducing in a self-consistent fashion appropriate transport gaps for both 
materials as determined from a combination of experimental and theoretical information. In these calculations we have also analyzed the metal-insulator phase-transition occurring in 
this interface when applying a bias voltage between the two materials. We find that at zero bias the interface is metallic while for bias larger than $\sim$ 0.5 eV (shifting TTF 
towards higher binding energies w.r.t. TCNQ) the system becomes an insulator. Our results have been favorably compared with the experimental data, indicating that the main mechanism 
controlling energy level alignment in the TTF/TCNQ interface is the charge transfer between both organic semiconductors. 

\section{Acknowledgements}

This work is supported by Spanish MICIIN under contract FIS2010-16046, the CAM under contract S2009/MAT-1467 and the European Project MINOTOR (Grant FP7-NMP-228424). JIB gratefully 
acknowledges financial support by the European Project MINOTOR. JIM acknowledges funding from Spanish MICIIN through Juan de la Cierva Program. Special thanks to Dr. Daniel Gonz\'alez 
for the development of the {\sc Xeo} software package, which has been extensively used for the data analysis. 



\end{document}